\pdfoutput=1

\documentclass{article}
\usepackage[preprint]{spconf}
\usepackage{amsmath,graphicx}
\usepackage{cite}
\usepackage{graphicx}
\usepackage{amsthm}      
\usepackage{amssymb}
\usepackage{dsfont}
\usepackage[ruled,vlined]{algorithm2e}
\usepackage{empheq}
\usepackage{float}
\usepackage{appendix}
\usepackage[skip=5pt]{caption}

\theoremstyle{plain}

\newtheoremstyle{mystyle}
  {}
  {}
  {\itshape}
  {}
  {\bfseries}
  {.}
  { }
  {}

\theoremstyle{mystyle}
\newtheorem{assumption}{Assumption}      
\newtheorem{theorem}{Theorem}

\title{Gramian-Based Adaptive Combination Policies\\for Diffusion Learning over Networks}

\name
 {Y. Efe Erginbas $^\star$, \thanks{$^\star$ Y. Efe Erginbas is with Department of Electrical and Electronics Engineering, Bilkent University, 06800 Bilkent, Ankara, Turkey. The author performed the work while at EPFL, Switzerland. E-mail: efe.erginbas@ug.bilkent.edu.tr}
 Stefan Vlaski $^\dagger$ and Ali H. Sayed $^\dagger$ \thanks{$^\dagger$ Stefan Vlaski and Ali H. Sayed are with School of Engineering, EPFL, CH-1015 Lausanne,
Switzerland. E-mails: \{stefan.vlaski,
    ali.sayed\}@epfl.ch}
 }
 \address{}

\begin{document}
\ninept

\setlength{\abovedisplayskip}{3pt}
\setlength{\belowdisplayskip}{2pt}
\setlength{\baselineskip}{1.1em}

\maketitle

\copyrightnotice{\begin{minipage}{\textwidth}\footnotesize\copyright\ 2020 IEEE. Personal use of this material is permitted. Permission from IEEE must be obtained for all other uses, in any current or future media, including reprinting/republishing this material for advertising or promotional purposes, creating new collective works, for resale or redistribution to servers or lists, or reuse of any copyrighted component of this work in other works.\end{minipage}}

\begin{abstract}
This paper presents an adaptive combination strategy for distributed learning over diffusion networks. Since learning relies on the collaborative processing of the stochastic information at the dispersed agents, the overall performance can be improved by designing combination policies that adjust the weights according to the quality of the data. Such policies are important because they would add a new degree of freedom and endow  multi-agent systems with the ability to control the flow of information over their edges for enhanced performance. Most adaptive and static policies available in the literature optimize certain performance metrics related to steady-state behavior, to the detriment of transient behavior. In contrast, we develop an adaptive combination rule that aims at optimizing the transient learning performance, while maintaining the enhanced steady-state performance obtained using policies previously developed in the literature.
\end{abstract}


\section{Introduction}

We consider a strongly-connected network of $N$ nodes with a predefined topology. We denote the neighborhood of node $k$ (including node $k$ itself) by $\mathcal{N}_k$ and the degree of node $k$ by $n_k$. A local risk function $J_k(w) = \mathds{E}_x \mathcal{Q}_k(w; \boldsymbol{x})$ is associated with each node, where $\mathcal{Q}_k(w; \boldsymbol{x})$ is some loss function. The objective is to generate an estimate, in a collaborative and distributed manner, for the unknown vector $w^o \in \mathbb{R}^M$ that minimizes the global cost:
\begin{equation}
    J^\mathrm{glob}(w) \triangleq \sum_{k = 1}^{N} J_k(w)
\end{equation}

\noindent
Each $J_k(w)$ is a real valued function defined over $w \in \mathbb{R}^M$, and assumed to be differentiable and strongly convex. Consequently, $J^\mathrm{glob}(w)$ is also a strongly convex and the minimizer $w^o$ is unique \cite{poliak_introduction_1987}. In this work, we focus on the important case where each of the local cost functions $\{J_k(w)\}$ are also minimized at the same $w^o$.

The solution to this problem can be pursued in a decentralized and iterative manner by generating local estimates $w_{k,i}$ at each node $k$ and time $i \geq 0$. The iterates can be constructed using a variety of decentralized strategies, including  incremental \cite{bertsekas_incremental_1996}, consensus \cite{nedic_dist_2009, nedic_constrained_2010, yuan_convergence_2013}, diffusion  \cite{sayed_book_2014, chen_diffusion_2012, chen_pareto_2013, chen_behavior_2015}, primal-dual \cite{towfic_2015}, proximal \cite{shi_proximal_2015}, augmented Lagrangian \cite{moura_linear_2015}, or gradient tracking methods \cite{kar_decentral_2020}. Here, we focus on the \textit{Adapt-then-Combine} (ATC) diffusion strategy, which has been shown to have superior mean-square error performance and stability range in adaptive scenarios \cite{chen_diffusion_2012}.

Since the statistical distribution of the data $\boldsymbol{x}$ is not known beforehand in most cases of interest, the exact gradient vectors $\nabla_w J_k(w)$ are not available or easily obtainable. Motivated by this consideration, we follow the stochastic gradient descent construction for diffusion algorithms and utilize gradient approximations in our analysis. We refer to the perturbation as a random additive noise component and write the approximate gradient vector in the form:
\begin{equation}
   \widehat{\nabla_w J}_k(w) = \nabla_w J_k(w) + \boldsymbol{s}_{k,i} (w)
   \label{eq:grad_noise}
\end{equation}

\noindent
where $\boldsymbol{s}_{k,i} (w)$ denotes the gradient noise term. Note that we use a boldface symbol to highlight its stochastic nature. Using the perturbed gradient vectors, the ATC diffusion strategy is given by:
\begin{subequations}
	\begin{empheq}[left=\empheqlbrace\,]{align}
	&\boldsymbol{\psi}_{k,i} = \boldsymbol{w}_{k,i-1} - \mu_k \widehat{\nabla_w J}_k(\boldsymbol{w}_{k,i-1}) \label{alg:ATC_random_a}\\
	&\boldsymbol{w}_{k,i} = \sum_{\ell \in \mathcal{N}_k} \boldsymbol{a}_{\ell k,i} \boldsymbol{\psi}_{\ell,i}  \label{alg:ATC_random_b}
	\end{empheq}
	\label{alg:ATC_random}
\end{subequations}

\noindent
where $\boldsymbol{\psi}_{k,i}$ is an intermediate local estimation vector at node $k$, $\mu_k$ is the step size at node $k$ and $\{\boldsymbol{a}_{\ell k,i}\}$ are possibly time-varying, real-valued combination weights. In most of the prior works on decentralized processing, the combination weights are pre-selected and treated as known deterministic variables. However, in this work, we focus on the case where these combination weights are constructed on the fly, and will therefore be data-dependent. They will nevertheless be always normalized to satisfy:
\begin{equation}
    \sum_{\ell \in \mathcal{N}_k} \boldsymbol{a}_{\ell k,i} = 1, \:\:\: \boldsymbol{a}_{\ell k,i} = 0 \text{ if } \ell \notin \mathcal{N}_k
    \label{eq:weight_constraint}
\end{equation}

\noindent
Although the combination weights are constrained to be non-negative in most previous works \cite{sayed_book_2014, chen_diffusion_2012, chen_pareto_2013, chen_behavior_2015}, the literature also includes works on diffusion strategies that allow negative values \cite{nassif_subspace_2019}. We will not impose the non-negativity requirement as well.

Since the choice of the combination weights $\{\boldsymbol{a}_{\ell k,i}\}$ plays an important role in the performance of decentralized strategies, different static and adaptive combination policies have been proposed in previous studies. The static combination policies include Uniform \cite{blondel_convergence_2005}, Laplacian \cite{scherber_locally_2004, xiao_fast_2004}, Maximum-Degree \cite{lin_xiao_scheme_2005}, Metropolis \cite{xiao_fast_2004}, Hastings\cite{zhao_performance_2012} and Relative-Variance \cite{zhao_imperfect_2012} rules. To allow for data-aware processing, adaptive combination policies have also been developed. These policies learn data statistics during the operation of the algorithm and adapt combination weights accordingly. Examples include the adaptive version of Relative-Variance rule \cite{zhao_imperfect_2012}, Phase-Switching algorithm \cite{fernandez_adjustment_2014}, and the policy proposed in \cite{takahashi_adaptive_2010}.

Previous works on adaptive combiners mostly concentrate on improving the steady-state performance of the nodes by incorporating the information obtained during the learning process \cite{zhao_imperfect_2012, fernandez_adjustment_2014}. However, these algorithms do not explicitly aim at improving the transient performance. Some empirical results indicate that they can be outperformed by simple static combination policies (such as the averaging rule) in the transient phase \cite{zhao_imperfect_2012, sayed_book_2014}. We address this problem in the context of networks with a general cost structure described previously, and propose an adaptive combination policy that will improve the transient performance of the decentralized learning algorithms while preserving the improved steady-state performance.

\section{Gramian-Based Adaptive Policy}

We first describe the error recursions corresponding to the ATC diffusion formulation in \eqref{alg:ATC_random} and start by defining the error vectors
\begin{empheq}{align}
&\widetilde{\boldsymbol{w}}_{k,i} = w^o - \boldsymbol{w}_{k,i} \text{\;,}
&&\widetilde{\boldsymbol{\psi}}_{k,i} = w^o - \boldsymbol{\psi}_{k,i}
\end{empheq}

\noindent
for all $k = 1,\dots,N$ and $i \geq 0$. Following the approach taken in \cite{sayed_book_2014}, we subtract both sides of equations \eqref{alg:ATC_random_a} and \eqref{alg:ATC_random_b} from $w^o$, substitute expression \eqref{eq:grad_noise} for the perturbed gradient vector, and call upon the mean-value theorem to express the error recursion as:
\begin{subequations}
	\begin{empheq}[left={\empheqlbrace\,}]{align}
	&\widetilde{\boldsymbol{\psi}}_{k,i} = [I_M - \mu_k \boldsymbol{H}_{k,i-1}]\widetilde{\boldsymbol{w}}_{k,i-1} + \mu_k \boldsymbol{s}_{k,i}(\boldsymbol{w}_{k,i-1})
	\label{alg:ATC_error_a}\\
	&\widetilde{\boldsymbol{w}}_{k,i} = \sum_{\ell \in \mathcal{N}_k} \boldsymbol{a}_{\ell k,i} \widetilde{\boldsymbol{\psi}}_{\ell,i}
	\label{alg:ATC_error_b}
	\end{empheq}
	\label{alg:ATC_error}
\end{subequations}

\noindent
where
\begin{equation}
    \boldsymbol{H}_{k,i-1} \triangleq \int_{0}^{1} \nabla_w^2 J_k(w^o - t \widetilde{\boldsymbol{w}}_{k,i-1}) dt
\end{equation}

\noindent
and $\nabla^2 J_k(\cdot)$ denotes the Hessian matrix of $J_k(\cdot)$. We also define the combination vectors and the combination matrix as follows:
\begin{equation}
\begin{aligned}
    \boldsymbol{a}_{k,i} & \triangleq \text{col}\{\boldsymbol{a}_{1k,i}, \dots, \boldsymbol{a}_{Nk,i}\} \in \mathbb{R}^N \text{ for } k = 1,\dots,N\\
    \boldsymbol{A}_{i} & \triangleq \left[ \boldsymbol{a}_{1,i},\dots, \boldsymbol{a}_{N,i}\right] \in \mathbb{R}^{N \times N}
\end{aligned}
\end{equation}

\subsection{Development of the Algorithm}

We define the problem of selecting the combination weights as an optimization problem where our goal is to maximize the improvement obtained in each time step by estimating the best possible combination weighting for each node. For this purpose, we introduce a performance metric for the network and formulate the optimization problem using this metric. We construct the network Square-Deviation measure as follows:
\begin{equation}
        \mathbf{SD}_{\mathrm{av}}(i) \triangleq \frac{1}{N} \|\widetilde{\boldsymbol{w}}_{i}\|^2 =  \frac{1}{N} \sum_{k=1}^{N}  \|\widetilde{\boldsymbol{w}}_{k,i}\|^2\\
        \label{eq:policy_cost}
\end{equation}

\noindent
Although it is similar to the widely-used MSD metric \cite{sayed_book_2014}, SD measures the error norm at each iteration $i$ in contrast to MSD that measures the expected norm of the steady-state error. It follows from the definition that $\text{MSD}_{\mathrm{av}} = \lim_{i \to \infty} \mathds{E}\left[\mathbf{SD}_{\mathrm{av}}(i)\right]$.

In order to account for error similarity between the nodes, we construct the Gramian matrix $\boldsymbol{Q}_i$ for the set of intermediate estimation error vectors $\{\widetilde{\boldsymbol{\psi}}_{k,i}\}$. Its entries are given by
\begin{equation}
    \left[\boldsymbol{Q}_i\right]_{k,\ell} \triangleq \widetilde{\boldsymbol{\psi}}_{k,i}^\mathsf{T} \widetilde{\boldsymbol{\psi}}_{\ell,i}
\end{equation}

\noindent or equivalently,
\begin{equation}
    \boldsymbol{Q}_i \triangleq \widetilde{\boldsymbol{\Psi}}_i^\mathsf{T} \widetilde{\boldsymbol{\Psi}}_i
\end{equation} 
\noindent
where $\widetilde{\boldsymbol{\Psi}}_i \triangleq \left[\widetilde{\boldsymbol{\psi}}_{1,i},\dots,\widetilde{\boldsymbol{\psi}}_{N,i} \right]$. Using these definitions, we can express \eqref{alg:ATC_error_b} as a matrix vector product:
\begin{equation}
    \widetilde{\boldsymbol{w}}_{k,i} = \widetilde{\boldsymbol{\Psi}}_i \boldsymbol{a}_{k,i}
    \label{eq:matrix_prod_eqv}
\end{equation}

\noindent
In order to pursue an optimal combination policy $\boldsymbol{A}_i$ as a function of the estimation errors $\widetilde{\boldsymbol{\psi}}_{k,i}$, we define an objective function using the $\mathrm{SD}$ measure defined in \eqref{eq:policy_cost} and substitute \eqref{eq:matrix_prod_eqv} to get
\begin{equation}
    \mathbf{SD}_{\mathrm{av}}(i) = \frac{1}{N} \sum_{k = 1}^{N} \|\boldsymbol{\widetilde{\Psi}}_i \boldsymbol{a}_{k,i}\|^2 = \frac{1}{N} \sum_{k = 1}^{N} \boldsymbol{a}_{k,i}^\mathsf{T}     \boldsymbol{Q}_i \boldsymbol{a}_{k,i}\\
\end{equation}

\noindent
Consequently, the problem of finding the combination weights that minimize the given error measure can be expressed as a constrained minimization problem in the form:
\begin{equation}
\begin{aligned}
    \min_{\{\boldsymbol{a}_{\ell k,i}\}}& &&\sum_{k = 1}^{N} \boldsymbol{a}_{k,i}^\mathsf{T}     \boldsymbol{Q}_i \boldsymbol{a}_{k,i}\\
    \textrm{s.t.}& &&\boldsymbol{a}_{\ell k,i} = 0 \textrm{, for all } \ell \notin \mathcal{N}_k \\
    & &&\boldsymbol{a}_{k,i}^\mathsf{T}\mathds{1} = 1 \text{, for } k = 1,\dots,N
    \label{eq:opt_coupled}
\end{aligned}
\end{equation}

\noindent
Since some entries of the vector $\boldsymbol{a}_{k,i}$ are constrained to be zero, we only need to solve this quadratic minimization problem for entries $\boldsymbol{a}_{\ell k,i}$ such that $\ell \in \mathcal{N}_k$. Therefore, we define the truncated combination vectors:
\begin{equation}
    \boldsymbol{c}_{k,i} \triangleq \text{col}\{\boldsymbol{a}_{\ell k,i}\}_{\ell \in \mathcal{N}_k} \in \mathbb{R}^{n_k}
\end{equation}

\noindent
such that 
\begin{equation}
    \boldsymbol{a}_{k,i} = P_k \boldsymbol{c}_{k,i}
    \label{eq:comb_eqv}
\end{equation}

\noindent
where $P_k \triangleq [\dots, \ell^\text{th} \text{ column of } I_N, \dots]$ for $\ell \in \mathcal{N}_k$ (i.e, $P_k$ is the $N \times n_k$ matrix whose columns are standard (natural) basis vectors of indices corresponding to the neighbors of node $k$). Additionally, we define the local counterparts of the Gramian matrix and error matrix as follows:
\begin{subequations}
    \begin{empheq}{align}
        \widetilde{\boldsymbol{\Psi}}_{k,i} &\triangleq \widetilde{\boldsymbol{\Psi}}_i P_k\\
        \boldsymbol{Q}_{k,i} &\triangleq \widetilde{\boldsymbol{\Psi}}_{k,i}^\mathsf{T} \widetilde{\boldsymbol{\Psi}}_{k,i} = P_k^\mathsf{T} \boldsymbol{Q}_{i} P_k
    \end{empheq}
\end{subequations}

\noindent
Furthermore, in \eqref{eq:opt_coupled}, both the $k^\text{th}$ term of the summation in the objective function and the constraints for $\boldsymbol{a}_{k,i}$ do not depend on the selection of $\boldsymbol{a}_{\ell,i}$ for $\ell \neq k$. Therefore, optimization problem  can be decoupled into $N$ independent sub-problems. Using the notation introduced in \eqref{eq:comb_eqv}, we express the sub-problem related to node $k$ as:
\begin{equation}
\begin{aligned}
    \min_{\boldsymbol{c}_{k,i}}& &&\boldsymbol{c}_{k,i}^\mathsf{T} \boldsymbol{Q}_{k,i} \boldsymbol{c}_{k,i}\\
    \textrm{s.t.}& &&\boldsymbol{c}_{k,i}^\mathsf{T}\mathds{1} = 1
\end{aligned}
\label{eq:min_seperate_truncated}
\end{equation}

\noindent
We continue with writing the optimality conditions as a KKT system for this equality constrained quadratic optimization problem:
\begin{equation}
    \begin{bmatrix}
    \boldsymbol{Q}_{k,i} & \mathds{1}_{n_k}\\
    \mathds{1}_{n_k}^\mathsf{T} & 0
    \end{bmatrix}
    \begin{bmatrix}
    \boldsymbol{c}_{k,i}\\
    \lambda
    \end{bmatrix}
    =
    \begin{bmatrix}
    0_{n_k}\\
    1
    \end{bmatrix}
    \label{eq:kkt_system}
\end{equation}

\noindent
For general quadratic optimization problems, if the KKT system is not solvable, it means that the problem is unbounded below or infeasible \cite[p. 522]{boyd_vandenberghe_2004}. However, $\boldsymbol{Q}_{k,i}$ is certainly positive-semidefinite (since it is a Gramian matrix) and thus the minimum value of the objective is bounded below by zero. Additionally, we note that the feasible set given by the linear constraint is non-empty. Therefore, solving \eqref{eq:kkt_system} will always yield an optimal $\boldsymbol{c}_{k,i}$ (there may be multiple optimal solutions if the KKT matrix is singular). 

If $\boldsymbol{Q}_{k,i}$ turns out to be non-singular, it is straightforward to verify that the optimal solution can be found in the following closed-form:
\begin{equation}
    \boldsymbol{c}_{k,i} = \frac{\boldsymbol{Q}_{k,i}^{-1}\mathds{1}_{n_k}} {\mathds{1}_{n_k}^\mathsf{T} \boldsymbol{Q}_{k,i}^{-1}\mathds{1}_{n_k}}
    \label{eq:comb_closed}
\end{equation}

\noindent
The resulting expressions enable us to compute the combination weights as a function of the matrix $\boldsymbol{Q}_{k,i}$ and hence $\widetilde{\boldsymbol{\psi}}_{k,i}$. However, these quantities are not known for the nodes, since the nodes only have access to $\boldsymbol{\psi}_{k,i}$, but not $w^o$. Consequently, the difficulty in employing the given combination policy, while optimal in the sense that it maximizes the reduction in squared deviation, lies in the fact that we are required to estimate the statistics of the error vectors.

In order to construct an estimate for the matrix $\boldsymbol{Q}_{k,i}$, we adopt a sample mean approach. We recall the definition of $\boldsymbol{Q}_{k,i}$ and express it in terms of the known quantities $\{\boldsymbol{\psi}_{k,i}\}$ as
\begin{equation}
    \boldsymbol{Q}_{k,i} \triangleq \widetilde{\boldsymbol{\Psi}}_{k,i}^\mathsf{T} \widetilde{\boldsymbol{\Psi}}_{k,i} = (\boldsymbol{\Psi}_{k,i} - w^o\mathds{1}_{n_k}^\mathsf{T})^\mathsf{T} (\boldsymbol{\Psi}_{k,i} - w^o\mathds{1}_{n_k}^\mathsf{T})
\end{equation}

\noindent
where $\boldsymbol{\Psi}_{k,i} \triangleq \left[\dots,\boldsymbol{\psi}_{\ell,i},\dots\right]$ for $\ell \in \mathcal{N}_k$. However, estimating $\boldsymbol{Q}_{k,i}$ directly in this form is difficult because it requires knowledge of the optimal weight vector $w^o$. We follow the reasoning used in \cite{zhao_imperfect_2012, takahashi_adaptive_2010}, and approximate the optimal $w^o$ by the expected estimate at iterate $i$, i.e., $ w^o \approx \mathds{E} \boldsymbol{\psi}_{k, i} $. Consequently, an estimate for $\boldsymbol{Q}_{k,i}$ will be
\begin{equation}
    \widehat{\boldsymbol{Q}}_{k,i} \approx (\boldsymbol{\Psi}_{k,i} - \mathbb{E}\boldsymbol{\Psi}_{k,i})^\mathsf{T} (\boldsymbol{\Psi}_{k,i} - \mathbb{E}\boldsymbol{\Psi}_{k,i})
    \label{eq:exp_avg_expression}
\end{equation}

\noindent
Using the approximation that $\boldsymbol{Q}_{k,i}$ is locally stationary, we propose using an exponential moving average scheme to construct $\widehat{\boldsymbol{Q}}_{k,i}$ as:
\begin{equation}
\begin{aligned}
    \widehat{\boldsymbol{Q}}_{k,i} &= (1-\alpha_1) \widehat{\boldsymbol{Q}}_{k,i-1} + \alpha_1 (\boldsymbol{\Psi}_{k,i} - \boldsymbol{\bar{\Psi}}_{k, i-1})^\mathsf{T} (\boldsymbol{\Psi}_{k,i} - \boldsymbol{\bar{\Psi}}_{k, i-1})\\
   \boldsymbol{\bar{\Psi}}_{k,i} &= (1-\alpha_2) \boldsymbol{\bar{\Psi}}_{k, i-1} + \alpha_2 \boldsymbol{\Psi}_{k,i}
   \label{eq:exp_avg_scheme}
\end{aligned}
\end{equation}

\noindent
for some constants $0 < \alpha_1, \alpha_2 \ll 1$. Using this estimation scheme, we can complete the construction of the algorithm as follows:
\begin{algorithm}
\caption{Gramian-Based Adaptive Diffusion}
\SetAlgoLined
 Parameters and initialization: $0 < \alpha_1, \alpha_2 \ll 1$, $\mu_k > 0$, 
 $\widehat{\boldsymbol{Q}}_{k,-1} = I_{n_k}$, $\boldsymbol{\bar{\Psi}}_{k,-1} = 0_{M\times n_k}$, $\boldsymbol{w}_{k,-1} = 0_M$.  \\
 \For{\text{each time } $i \geq 0$}{
    \For{each node $k$}{
  	$\boldsymbol{\psi}_{k,i} = \boldsymbol{w}_{k,i-1} - \mu_k \widehat{\nabla_w J}_k(\boldsymbol{w}_{k,i-1})$\\
	}
    \For{each node $k$}{
        $\boldsymbol{\Psi}_{k,i} =\left[\dots, \boldsymbol{\psi}_{\ell,i}, \dots \right]$ , for $\ell \in \mathcal{N}_k$\\
        $\boldsymbol{G}_{k,i} = (\boldsymbol{\Psi}_{k,i} - \boldsymbol{\bar{\Psi}}_{k,i-1})^\mathsf{T} (\boldsymbol{\Psi}_{k,i} - \boldsymbol{\bar{\Psi}}_{k,i-1})$\\
        $\widehat{\boldsymbol{Q}}_{k,i} = (1-\alpha_1) \widehat{\boldsymbol{Q}}_{k,i-1} + \alpha_1 \boldsymbol{G}_{k,i}$ \\
        $\boldsymbol{\bar{\Psi}}_{k,i} = (1-\alpha_2) \boldsymbol{\bar{\Psi}}_{k,i-1} + \alpha_2 \boldsymbol{\Psi}_{k,i}$\\
        solve $\begin{bmatrix}
        \widehat{\boldsymbol{Q}}_{k,i} & \mathds{1}_{n_k}\\
        \mathds{1}_{n_k}^\mathsf{T} & 0
        \end{bmatrix}
        \begin{bmatrix}
        \boldsymbol{c}_{k,i}\\
        \lambda
        \end{bmatrix}
        =
        \begin{bmatrix}
        0_{n_k}\\
        1
        \end{bmatrix}$ , for $\boldsymbol{c}_{k,i}$\\
    	$\boldsymbol{w}_{k,i} = \boldsymbol{\Psi}_{k,i} \boldsymbol{c}_{k,i}$
    }
 }
\end{algorithm}

\vspace{-5mm}
\section{Steady-State Mean-Square Performance}

To examine the performance of the algorithm, it is necessary to introduce some simplifying assumptions; otherwise, the analysis becomes intractable due to the multiple adaptation layers. Some of these assumptions are common in the literature on decentralized adaptive algorithms. They essentially essentially require that the construction of the approximate gradient vector should not introduce bias and that its error variance should decrease as the quality of the iterate improves \cite{sayed_book_2014}.

\begin{assumption}[\bfseries Conditions on Gradient Noise]
\label{assumption:grad_noise}

The gradient noise processes are temporally and spatially independent. Additionally, the first and second-order moments of the gradient noise processes satisfy the following conditions:
\begin{subequations}
    \begin{empheq}{align}
     \mathbb{E} \left[ \boldsymbol{s}_{k,i}(\boldsymbol{w}_{k,i-1}) \right] &= 0\\
    \mathbb{E} \left[ \| \boldsymbol{s}_{k,i}(\boldsymbol{w}_{k,i-1}) \|^2 \right] &\leq \beta^2 \mathbb{E}\left[ \| w^o - \boldsymbol{w}_{k,i-1} \|^2 \right]  + \sigma_{s,k}^2
    \end{empheq}
\end{subequations}

\noindent
for some constants $\beta^2 \geq 0$, $\sigma_{s,k}^2 \geq 0$.
\qed

\end{assumption}

\noindent
For static combination matrices that satisfy \eqref{eq:weight_constraint}, it has been shown that $\|\widetilde{\boldsymbol{w}}_{k,i}\|^2$ can be made arbitrarily small in steady-state if sufficiently small step-sizes are used \cite{sayed_book_2014, chen_diffusion_2012}. Consequently, it has been argued that $\boldsymbol{H}_{k,i-1}$ can be approximated by $H_k^o \triangleq \nabla_w^2 J_k(w^o)$ for small step-sizes. In light of this observation, we employ the assumption that similar arguments will hold for our dynamic combination policy. In other words, we argue that the iterates can get sufficiently close to $w^o$ such that the curvature and noise structures can be well-approximated with their corresponding values at $w^o$. 

Furthermore, as done in the analysis of other adaptive policies in the literature \cite{takahashi_adaptive_2010, zhao_imperfect_2012}, we will assume that the estimation process is wide-sense stationary and samples are temporally uncorrelated in steady-state. This follows from the observation that the estimates $\boldsymbol{w}_{k,i}$ will approach $w^o$ in expectation and the deviations will be mostly caused by independent noise factors.

\begin{assumption}[\bfseries Steady-State]
Using small enough step-sizes, it is assumed that the following stationarity conditions are approximately valid:
\begin{subequations}
    \begin{empheq}{align}
        \lim_{i \to \infty} \boldsymbol{H}_{k,i-1} &\approx H_k^o\\
        \lim_{i \to \infty} \mathbb{E} \left[ \boldsymbol{s}_{k,i}(\boldsymbol{w}_{k,i-1})  \boldsymbol{s}_{k,i}^\mathsf{T}(\boldsymbol{w}_{k,i-1}) \right] &\approx R_{s,k}\\
        \lim_{i \to \infty} \mathbb{E}\boldsymbol{\Psi}_i &\approx \mathbb{E}\boldsymbol{\Psi}\\
        \lim_{i \to \infty} \mathbb{E}\left[ \boldsymbol{\Psi}_{i}^\mathsf{T} \boldsymbol{\Psi}_{i-1} \right] &\approx \mathbb{E}\boldsymbol{\Psi}^\mathsf{T} \mathbb{E}\boldsymbol{\Psi}
    \end{empheq}
\end{subequations}
\noindent
where $H_k^o \triangleq \nabla_w^2 J_k(w^o)$, $R_{s,k} \triangleq \mathbb{E}\left[ \boldsymbol{s}_{k,i}(w^o) \boldsymbol{s}_{k,i}^\mathsf{T}(w^o) \right]$ and $\mathbb{E}\boldsymbol{\Psi}$ is an unknown deterministic matrix.
\qed

\label{assumption:steady_state}
\end{assumption}

\noindent
Lastly, we will require the independence of each combination matrix from the last estimation error. For small enough $\alpha_1$ and $\alpha_2$ values, the combination matrix $\boldsymbol{A}_i$ at step $i$ is a function of many past noise samples, while the estimation error $\widetilde{\boldsymbol{w}}_{i-1}$ at step $i$ mostly depends on more recent noise samples. Therefore, it is reasonable to employ the assumption that two quantities are independent.

\begin{assumption}[\bfseries Independent Combination Matrix]
For small $\alpha_1$ and $\alpha_2$ values, $\boldsymbol{A}_i$ and $\widetilde{\boldsymbol{w}}_{k,i-1}$ are independent in steady-state.
\label{assumption:w_F_ind}
\end{assumption}

\subsection{Approximate Steady-State Combination Policy} \label{subsec:apprx}

Since the expression for the combination policy and the distribution of $\boldsymbol{\Psi}_i$ matrices are intertwined with each other, it is not straightforward to write down a steady-state expression for $\mathbb{E} \boldsymbol{A}_{i}$. However, we can introduce the assumption that $\mathbb{E} \boldsymbol{A}_{i}$ will converge to some constant matrix, following \cite{takahashi_adaptive_2010}. This assumption originates from the observation that $\boldsymbol{A}_{i}$ matrices change slowly over iterations (for small $\alpha_1$ and $\alpha_2$ values) and therefore capture information mostly about the error statistics of the nodes. Furthermore, we assume that this constant matrix is such that it minimizes the expected error at the nodes and hence it is equal to the combination matrix generated by \eqref{eq:comb_closed} using $\lim_{i \to \infty} \mathbb{E}\widehat{\boldsymbol{Q}}_{k,i}$ for $\boldsymbol{Q}_{k,i}$. Essentially, we assume that the expectation applied to both sides of \eqref{eq:comb_closed} can be approximated as follows in the steady-state limit:
\begin{equation}
    \mathbb{E}\boldsymbol{c}_{k,i} = \mathbb{E} \left[ \frac{\boldsymbol{Q}_{k,i}^{-1}\mathds{1}_{n_k}} {\mathds{1}_{n_k}^\mathsf{T} \boldsymbol{Q}_{k,i}^{-1}\mathds{1}_{n_k}} \right] \approx \frac{\mathbb{E}\left[\boldsymbol{Q}_{k,i}\right]^{-1}\mathds{1}_{n_k}} {\mathds{1}_{n_k}^\mathsf{T} \mathbb{E}\left[\boldsymbol{Q}_{k,i}\right]^{-1}\mathds{1}_{n_k}}
\end{equation}

\begin{assumption}
Expected value of the combination matrix $\boldsymbol{A}_{i}$ converges to $A_{\infty}$ in steady-state regime and we can employ following approximations:
\begin{subequations}
    \begin{empheq}{align}
    \lim_{i \to \infty} \mathbb{E}\left[ \boldsymbol{A}_{i} \right] &\approx A_{\infty}\\
    \lim_{i \to \infty} \mathbb{E}\left[ \boldsymbol{\mathcal{A}}_{i} \otimes \boldsymbol{\mathcal{A}}_{i}\right] &\approx \mathcal{A}_{\infty} \otimes \mathcal{A}_{\infty}
    \end{empheq}
\end{subequations}

\noindent
where $\boldsymbol{\mathcal{A}}_{i} = \boldsymbol{A}_{i} \otimes I_M$, $\mathcal{A}_{\infty} = A_{\infty} \otimes I_M$ and columns of $A_{\infty}$ are
\begin{equation}
    a_{k,\infty} = \frac{P_k \widehat{Q}_{k,\infty}^{-1}\mathds{1}_{n_k}} {\mathds{1}_{n_k}^\mathsf{T} \widehat{Q}_{k,\infty}^{-1}\mathds{1}_{n_k}}
    \label{eq:a_k_inf}
\end{equation}
\label{assumption:limit_policy}
\noindent
where $\widehat{Q}_{k,\infty} \triangleq P_k^T \widehat{Q}_{\infty} P_k$ and $\widehat{Q}_{\infty} \triangleq \lim_{i \to \infty} \mathbb{E}\widehat{\boldsymbol{Q}}_{i}$.
\qed
\end{assumption}

\noindent
Following this assumption, we write an expression for $\mathbb{E}\widehat{\boldsymbol{Q}}_{i}$ in steady-state, so that we can approximate $A_\infty$.

\newpage
\begin{theorem}
\label{THM:EXP_Q}
Under Assumptions \ref{assumption:grad_noise}-\ref{assumption:steady_state}, the steady state expectation for $\widehat{\boldsymbol{Q}}_{i}$ can be approximated as a diagonal matrix of the form:
\begin{equation}
    \lim_{i \to \infty} \mathbb{E}\widehat{\boldsymbol{Q}}_{i} \approx \frac{1}{2} \text{diag} \{\mu_1^2 \sigma_{s,1}^2, \dots, \mu_N^2 \sigma_{s,N}^2\} 
     \label{eq:Q_inf}
\end{equation}

\noindent
where $\sigma_{s,k}^2 \triangleq \mathbb{E} \| \boldsymbol{s}_{k,i}(w^o) \|^2$.

\end{theorem}

\begin{proof}
The proof is omitted due to space limitations. 
\end{proof}

\noindent
When we substitute \eqref{eq:Q_inf} into \eqref{eq:a_k_inf}, each entry of the matrix $A_{\infty}$ is found to be approximated by
\begin{equation}
    a_{\ell k,\infty} = 
    \begin{cases}
    \begin{aligned}
    &\frac{\theta_\ell}{ \sum_{m \in \mathcal{N}_k} \theta_m } && \text{, if } \ell \in \mathcal{N}_k\\
    &0 && \text{, otherwise}
    \end{aligned}
    \end{cases}
    \label{eq:A_entries}
\end{equation}

\noindent
where $\theta_\ell \triangleq 1/(\mu_\ell^2\sigma_{s,\ell}^2)$. As this result shows, the steady-state combination weights generated by the proposed algorithm matches in expectation with the Relative-Variance rule \cite{sayed_book_2014, zhao_imperfect_2012}. Of course, this result only holds under the simplifying Assumption \ref{assumption:limit_policy}. Nevertheless, as we illustrate numerically in Sec. \ref{sec:sim}, the resulting approximation error is small in practice. Therefore, we can conclude that our proposed algorithm can match the enhanced steady-state performance of previously proposed algorithms.

\subsection{Steady-State Mean-Square Performance}

Our next goal is to approximate the steady-state network MSD. We follow the analysis conducted in \cite{sayed_book_2014, chen_diffusion_2012} and adapt it according to our case of study. We also use the low-rank approximation method described in \cite{sayed_book_2014} so that network MSD can be approximated in terms of the data/noise statistics, the Perron eigenvector of the combination policy, and the step sizes.

\begin{theorem}[\bfseries Low-Rank Approximation for the Network MSD]
\label{THM:APPRX_MSD}
Under Assumptions \ref{assumption:grad_noise}-\ref{assumption:limit_policy}, the steady-state network MSD obtained by the proposed algorithm is approximately equal to
\begin{equation}
\mathrm{MSD}_{\mathrm{av}} \approx \frac{1}{2} \mathrm{Tr} \left[ \left(\sum_{k=1}^{N} \mu_k p_k H_k^o\right)^{-1} \left(\sum_{k=1}^{N} \mu_k^2 p_k^2 R_{s,k}\right) \right]
\label{eq:low_rank_msd}
\end{equation}

\noindent
where $ H_k^o = \nabla_w^2 J_k(w^o)$, $R_{s,k} = \mathbb{E}\left[ \boldsymbol{s}_{k,i}(w^o) \boldsymbol{s}_{k,i}^\mathsf{T}(w^o) \right]$ and
\begin{equation}
    p_k = \frac{ \theta_k \sum_{m \in \mathcal{N}_k}  \theta_m  }{ \sum_{\ell=1}^{N} \left(\theta_\ell  \sum_{m \in \mathcal{N}_\ell}  \theta_m \right) }
    \label{eq:perron_entries}
\end{equation}

\noindent
using the notation $\theta_k = 1/(\mu_k^2\sigma_{s,k}^2)$ , for $k = 1,\dots,N$.

\end{theorem}

\begin{proof}
The proof is omitted due to space limitations. 
\end{proof}

\section{Simplified Version of the Algorithm}

In the steady-state analysis of the originally proposed algorithm, we have observed that the expected value of the matrix $\boldsymbol{Q}_i$ converges to a diagonal matrix as $i$ goes to infinity. Since computation of only diagonal elements would also result in the same steady-state MSD with our original approach, we consider a diagonal approximation for $\boldsymbol{Q}_i$ starting from the initial iterate. Therefore, one approximation scheme for $\boldsymbol{Q}_i$ can be expressed as
\begin{equation}
\begin{aligned}
    \boldsymbol{q}_{k,i} &= (1-\alpha_1) \boldsymbol{q}_{k,i-1} + \alpha_1 \| \boldsymbol{\psi}_{k,i} - \bar{\boldsymbol{\psi}}_{k,i-1} \|^2 \\
   \bar{\boldsymbol{\psi}}_{k,i} &= (1-\alpha_2) \bar{\boldsymbol{\psi}}_{k,i-1} + \alpha_2 \boldsymbol{\psi}_{k,i}
    \label{eq:Q_only_diag}
\end{aligned}
\end{equation}
\noindent
where $\boldsymbol{q}_{k,i}$ is the $k^{th}$ diagonal entry of the approximation $\widehat{\boldsymbol{Q}}_i$. Using \eqref{eq:comb_closed}, each entry of the matrix $\boldsymbol{A}_{i}$ becomes equal to
\begin{equation}
    \boldsymbol{a}_{\ell k,i} = 
    \begin{cases}
    \begin{aligned}
    &\frac{\boldsymbol{q}_{\ell,i}^{-1}}{ \sum_{m \in \mathcal{N}_k} \boldsymbol{q}_{m,i}^{-1} } && \text{ if } \ell \in \mathcal{N}_k\\
    &0 && \text{ otherwise}
    \end{aligned}
    \end{cases}
    \label{eq:A_entries_simpl}
\end{equation}

\begin{figure}
    \centering
    \includegraphics[width=\columnwidth]{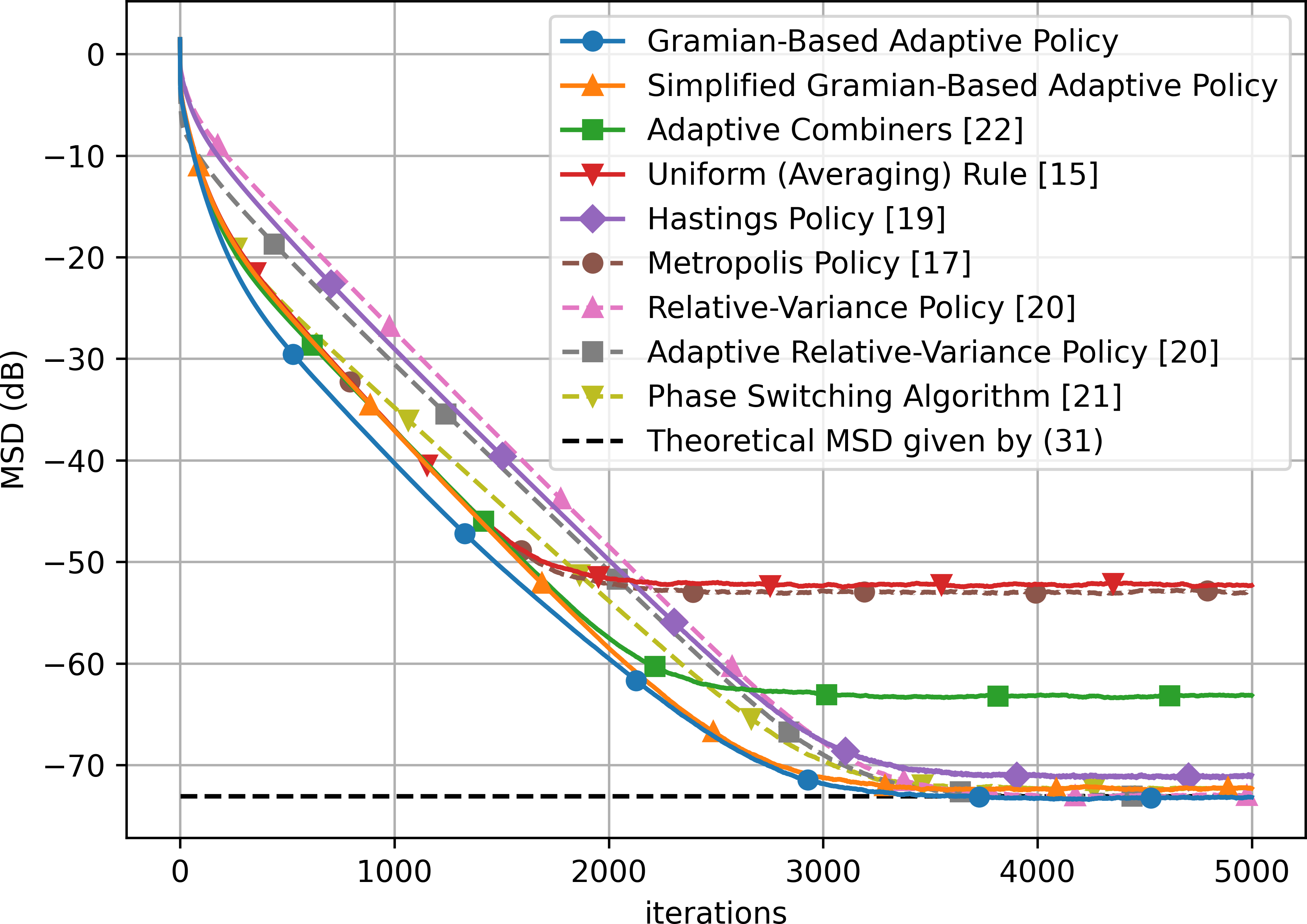}
    \caption{Learning curves obtained with different combination policies}
    \label{fig:experiment_res_log}
    \vspace{-3mm}
\end{figure}

\section{Simulations}
\vspace{-1mm}
\label{sec:sim}

We provide an empirical performance analysis of various combination policies with ATC algorithm on a decentralized logistic regression problem. The objective is to generate estimates for optimal $w^o$ that minimizes all of the local cost functions:
\begin{equation}
    J_k(w) = \mathbb{E} \left\{ \text{ln} (1+e^{-\boldsymbol{\gamma}_k \boldsymbol{h}_k^\mathsf{T} w} ) \right\} + \frac{\rho_k}{2} \|w\|^2
\end{equation}

\noindent
where $(\boldsymbol{h}_k, \boldsymbol{\gamma}_k)$ represent the streaming data received at node $k$. We consider the stochastic construction given by instantaneous data samples $(\boldsymbol{h}_{k,i}, \boldsymbol{\gamma}_{k,i})$. At each iteration $i$, we first generate labels $\boldsymbol{\gamma}_{k,i} \in \{\pm 1\}$ independently and then generate the corresponding feature vector $\boldsymbol{h}_{k,i} \sim \mathcal{N}(\boldsymbol{\gamma}_{k,i}  \mu_{h,k}\mathds{1}_M, \sigma_{h,k}^2 I_M) \in \mathbb{R}^M$. In all experiments, we set $M = 10$, $N=20$, $\mu_k = 0.005$ and select $\rho_k$ approximately equal to $0.5$ for all nodes, where the minor adjustments serve the purpose of ensuring a common minimizer among all local cost functions.  Scaling factors $\mu_{h,k}$ and $\sigma_{h,k}^2$ are selected uniformly in the interval $(0.6,1.4)$ and log-uniformly in the interval $(10^{-2},1)$, respectively. All other algorithm related parameters are selected such that the observed average network MSD is minimized. For both versions of the proposed algorithm, the parameters are set to $\alpha_1 = 0.01$ and $\alpha_2 = 0.03$.

The plots in Fig.\ref{fig:experiment_res_log} depict the expected value of the network SD, which is approximated by averaging the results obtained from 400 independent experiments that use the same data statistics. Furthermore, we numerically approximate $H_k^o$ and $R_{s,k}$ matrices using the expressions given in \cite[p. 321]{sayed_book_2014}, so that we can calculate the theoretical MSD given by equation \eqref{eq:low_rank_msd}. We observe that the MSD obtained using either version of the Gramian-Based Adaptive Diffusion strategy agrees with the theoretical value and it also coincides with the results obtained by the Relative-Variance rule as discussed in Sect. \ref{subsec:apprx}. Proposed strategies maintain the best steady-state performance achieved by other combination rules. Moreover, both versions of the proposed algorithm outperform all others during the transient-phase, satisfying our primary objective in their development.

\vspace{-1mm}
\section{Conclusion}
\vspace{-1mm}

We proposed an adaptive combination rule for distributed estimation over diffusion networks. We achieved this formulation by defining an optimization problem where the goal is to obtain the least estimation error possible in each time step. We analyzed the performance of the proposed algorithm and concluded that it can maintain the enhanced MSD performance provided by previous algorithms in the literature, while improving the transient performance.

\newpage

\bibliographystyle{IEEEbib}
\bibliography{adaptive_combiners}

\begin{thebibliography}{10}

\bibitem{poliak_introduction_1987}
B.~T. Polyak,
\newblock {\em Introduction to Optimization},
\newblock Optimization Software, Publications Division, New York, 1987.

\bibitem{bertsekas_incremental_1996}
D.~P. Bertsekas,
\newblock ``A new class of incremental gradient methods for least squares
  problems,''
\newblock {\em SIAM Journal on Optimization}, vol. 7, pp. 913--926, Apr. 1996.

\bibitem{nedic_dist_2009}
A.~Nedic and A.~Ozdaglar,
\newblock ``Distributed subgradient methods for multi-agent optimization,''
\newblock {\em IEEE Transactions on Automatic Control}, vol. 54, no. 1, pp.
  48--61, Jan. 2009.

\bibitem{nedic_constrained_2010}
A.~{Nedic}, A.~{Ozdaglar}, and P.~A. {Parrilo},
\newblock ``Constrained consensus and optimization in multi-agent networks,''
\newblock {\em IEEE Transactions on Automatic Control}, vol. 55, no. 4, pp.
  922--938, Feb. 2010.

\bibitem{yuan_convergence_2013}
K.~Yuan, Q.~Ling, and W.~Yin,
\newblock ``On the convergence of decentralized gradient descent,''
\newblock {\em SIAM Journal on Optimization}, vol. 26, no. 3, pp. 1835 -- 1854,
  Oct. 2013.

\bibitem{sayed_book_2014}
A.~H. Sayed,
\newblock ``Adaptation, learning, and optimization over networks,''
\newblock {\em Foundations and Trends in Machine Learning}, vol. 7, no. 4-5,
  pp. 311--801, July 2014.

\bibitem{chen_diffusion_2012}
J.~Chen and A.~H. Sayed,
\newblock ``Diffusion {adaptation} {strategies} for {distributed}
  {optimization} and {learning} {over} {networks},''
\newblock {\em IEEE Transactions on Signal Processing}, vol. 60, no. 8, pp.
  4289--4305, Aug. 2012.

\bibitem{chen_pareto_2013}
J.~{Chen} and A.~H. {Sayed},
\newblock ``Distributed {Pareto} optimization via diffusion strategies,''
\newblock {\em IEEE Journal of Selected Topics in Signal Processing}, vol. 7,
  no. 2, pp. 205--220, Feb. 2013.

\bibitem{chen_behavior_2015}
J.~{Chen} and A.~H. {Sayed},
\newblock ``On the learning behavior of adaptive networks - part {I}: Transient
  analysis,''
\newblock {\em IEEE Transactions on Information Theory}, vol. 61, no. 6, pp.
  3487--3517, Apr. 2015.

\bibitem{towfic_2015}
Z.~J. Towfic and A.~H. Sayed,
\newblock ``Stability and performance limits of adaptive primal-dual
  networks,''
\newblock {\em IEEE Transactions on Signal Processing}, vol. 63, pp.
  2888--2903, June 2015.

\bibitem{shi_proximal_2015}
W.~{Shi}, Q.~{Ling}, G.~{Wu}, and W.~{Yin},
\newblock ``A proximal gradient algorithm for decentralized composite
  optimization,''
\newblock {\em IEEE Transactions on Signal Processing}, vol. 63, no. 22, pp.
  6013--6023, Nov. 2015.

\bibitem{moura_linear_2015}
D.~{Jakovetić}, J.~M.~F. {Moura}, and J.~{Xavier},
\newblock ``Linear convergence rate of a class of distributed augmented
  lagrangian algorithms,''
\newblock {\em IEEE Transactions on Automatic Control}, vol. 60, no. 4, pp.
  922--936, Apr. 2015.

\bibitem{kar_decentral_2020}
R.~{Xin}, S.~{Kar}, and U.~A. {Khan},
\newblock ``Decentralized stochastic optimization and machine learning: A
  unified variance-reduction framework for robust performance and fast
  convergence,''
\newblock {\em IEEE Signal Processing Magazine}, vol. 37, no. 3, pp. 102--113,
  May 2020.

\bibitem{nassif_subspace_2019}
R.~{Nassif}, S.~{Vlaski}, and A.~H. {Sayed},
\newblock ``Distributed inference over networks under subspace constraints,''
\newblock {\em IEEE ICASSP}, pp. 5232--5236, May 2019.

\bibitem{blondel_convergence_2005}
V.~D. {Blondel}, J.~M. {Hendrickx}, A.~{Olshevsky}, and J.~N. {Tsitsiklis},
\newblock ``Convergence in multiagent coordination, consensus, and flocking,''
\newblock {\em IEEE Conference on Decision and Control}, pp. 2996--3000, Dec.
  2005.

\bibitem{scherber_locally_2004}
D.~S. Scherber and H.~C. Papadopoulos,
\newblock ``Locally constructed algorithms for distributed computations in
  ad-hoc networks,''
\newblock {\em International Symposium on {Information} Processing in Sensor
  Networks}, pp. 11--19, Apr. 2004.

\bibitem{xiao_fast_2004}
L.~{Xiao} and S.~{Boyd},
\newblock ``Fast linear iterations for distributed averaging,''
\newblock {\em IEEE International Conference on Decision and Control}, vol. 5,
  pp. 4997--5002, Dec. 2003.

\bibitem{lin_xiao_scheme_2005}
L.~Xiao, S.~Boyd, and S.~Lall,
\newblock ``A scheme for robust distributed sensor fusion based on average
  consensus,''
\newblock {\em {International} {Symposium} on {Information} {Processing} in
  {Sensor} {Networks}}, pp. 63--70, Apr. 2005.

\bibitem{zhao_performance_2012}
X.~Zhao and A.~H. Sayed,
\newblock ``Performance {limits} for {distributed} {estimation} {over} {LMS}
  {adaptive} {networks},''
\newblock {\em IEEE Transactions on Signal Processing}, vol. 60, no. 10, pp.
  5107--5124, Oct. 2012.

\bibitem{zhao_imperfect_2012}
X.~Zhao, S-Y Tu, and A.~H. Sayed,
\newblock ``Diffusion adaptation over networks under imperfect information
  exchange and non-stationary data,''
\newblock {\em IEEE Transactions on Signal Processing}, vol. 60, no. 7, pp.
  3460--3475, July 2012.

\bibitem{fernandez_adjustment_2014}
J.~Fernandez-Bes, J.~Arenas-Garcia, and A.~H. Sayed,
\newblock ``Adjustment of combination weights over adaptive diffusion
  networks,''
\newblock {\em {IEEE} {ICASSP}}, pp. 6409--6413, May 2014.

\bibitem{takahashi_adaptive_2010}
N.~Takahashi, I.~Yamada, and A.~H. Sayed,
\newblock ``Diffusion least-mean squares with adaptive combiners: Formulation
  and performance analysis,''
\newblock {\em IEEE Transactions on Signal Processing}, vol. 58, no. 9, pp.
  4795--4810, June 2010.

\bibitem{boyd_vandenberghe_2004}
S.~Boyd and L.~Vandenberghe,
\newblock {\em Convex Optimization},
\newblock Cambridge University Press, Cambridge, 2004.

\end{thebibliography}

\end{document}